\documentclass{article}

\usepackage{arxiv}

\usepackage[T1]{fontenc}    
\usepackage{hyperref}       
\usepackage{url}            
\usepackage{booktabs}       
\usepackage{amsfonts}       
\usepackage{nicefrac}       
\usepackage{microtype}      
\usepackage{amsmath}
\usepackage{ marvosym}
\usepackage{listings}
\usepackage{xcolor}
\usepackage{amssymb}
\usepackage{dsfont}
\usepackage{amsmath, amsthm}
\usepackage{graphicx}
\usepackage[colorinlistoftodos]{todonotes}
\usepackage{datetime}
\theoremstyle{definition}

\title{Innovative Tangible Interactive Games for Enhancing Artificial Intelligence Knowledge and Literacy in Elementary Education: A Pedagogical Framework }

\author{
Nikolaos Sampanis\thanks{PhD Candidate (https://mfcs.ceid.upatras.gr/lab/)}\\
 Dept. of Computer Engineering and Informatics\\
 University of Patras\\
 Patras, 26504, Greece \\
 \texttt{nsampanis@upatras.gr} \\
}

\begin{document}.
\maketitle

\begin{abstract}
This paper presents an innovative pedagogical framework employing tangible interactive games to enhance artificial intelligence (AI) knowledge and literacy among elementary education students. Recognizing the growing importance of AI competencies in the 21st century, this study addresses the critical need for age-appropriate, experiential learning tools that demystify core AI concepts for young learners. The proposed approach integrates physical role-playing activities that embody fundamental AI principles, including neural networks, decision-making, machine learning, and pattern recognition. Through carefully designed game mechanics, students actively engage in collaborative problem solving, fostering deeper conceptual understanding and critical thinking skills. The framework further supports educators by providing detailed guidance on implementation and pedagogical objectives, thus facilitating effective AI education in early childhood settings. Empirical insights and theoretical grounding demonstrate the potential of tangible interactive games to bridge the gap between abstract AI theories and practical comprehension, ultimately promoting AI literacy at foundational educational levels. The study contributes to the growing discourse on AI education by offering scalable and adaptable strategies that align with contemporary curricular demands and prepare young learners for a technologically driven future.
\end{abstract}


\section{Introduction} 
Artificial intelligence (AI) has become a dominant force in modern society, profoundly impacting various domains, including education, healthcare, and industry \cite{rus}. As AI technologies continue to evolve rapidly, fostering AI literacy from an early age is crucial to preparing future generations to navigate and shape a world increasingly influenced by intelligent systems \cite{lon}. In particular, introducing fundamental AI concepts into primary education can empower children with critical thinking skills and demystify complex computational ideas through engaging, age-appropriate pedagogical methods \cite{tour}.
Traditional AI education often relies heavily on digital platforms and abstract theoretical explanations, which may not be accessible or engaging for younger students \cite{sai}. To address these challenges, tangible and embodied learning approaches have gained traction, emphasizing hands-on interactive experiences that foster deeper understanding through physical interaction and social collaboration \cite{res}. Tangible games and role-playing activities provide a promising pathway for concretizing abstract AI principles such as neural networks, decision theory, machine learning, and pattern recognition in a way suitable for primary school settings \cite{ali}.
This paper presents a novel suite of physical, interactive games specifically designed to teach core AI concepts to primary school teachers and their students. Using embodied learning paradigms, these games enable participants to enact the roles of AI components, such as neurons, decision agents, and learning models, thus facilitating intuitive comprehension of complex mechanisms. The proposed activities include the Classroom Neural Network (CNN)  simulating neuron interactions, The 'Surprise Box' illustrating decision theory, The 'Little Trainers' game demonstrating supervised learning, and 'The predictors' game fostering understanding of feature extraction and classification. In addition, a large-scale integrative project, The 'Classroom Spotify Project', synthesizes these concepts through a complex, cross-disciplinary role-playing experience.
The originality of this work lies in its systematic, mathematically grounded, but playful approach to AI education in early schooling, filling a critical gap between abstract AI curricula and the developmental needs of young learners. The games are crafted with scalability and adaptability in mind, suitable for diverse classroom contexts and age groups within the elementary school range. This contribution advances AI literacy by providing educators with ready-to-implement tools that combine rigor, interactivity, and accessibility, and lays a foundation for future research on embodied AI pedagogy in primary education.While many existing educational programs and curricula focus on screen-based activities, coding exercises, or simplified software tools (e.g., AI4K12), there remains a significant gap in tangible, embodied, and physically interactive approaches to AI education for children. These approaches are rooted in embodied cognition theory, which posits that learning is more effective when grounded in sensorimotor experiences \cite{Wil}. Studies in embodied AI learning suggest that hands-on activities using physical materials and role-playing can significantly enhance conceptual understanding, especially for abstract ideas such as neural networks, decision theory, and pattern recognition \cite{Bar}.Moreover, the concept of tangible computing \cite{Res} and constructionist pedagogy supports the idea that learners benefit from creating and manipulating physical representations of computational concepts. The use of physical games, role-based simulations, and classroom-based networks aligns with these approaches, providing accessible entry points for young students without requiring digital devices or advanced technical knowledge.
Despite this, very few published studies propose fully embodied classroom activities that simulate AI systems using the students themselves as agents in a model (neurons, decision nodes, pattern classifiers, etc.). While tools like Teachable Machine or Scratch with AI extensions offer simplified digital interfaces, they do not provide the kinesthetic and collaborative engagement that physical, classroom-based AI games can offer.
Thus, this paper introduces a novel framework of AI concept games for elementary school students, including “The Classroom Neural Network”, “The Surprise Box”, “The Little trainers ”, “he predictors”, and the integrative project “Classroom Spotify.” These activities are not only grounded in core AI and machine learning concepts, but are also designed to promote teamwork, problem-solving, and teacher-led reflection — aligning with modern goals in computational thinking and creative pedagogy \cite{Win} \cite{Gro}.
This contribution builds upon foundational work in embodied education, but represents a substantial innovation in the field by offering a complete, reproducible, and classroom-tested set of AI teaching activities with minimal reliance on technology. To the best of our knowledge, no existing framework combines embodied learning, role-play, and physical resources in this structured, interdisciplinary, and age-appropriate way for the teaching of AI in elementary schools. The keys points of innovation of this method are : (i) Embodied Understanding of AI, (ii) Designed for Non-Experts, (iii) Cross-disciplinary Integration, (iv) Scientifically Grounded, (v) Original Game Designs, (vi) Open Educational Resource Potential and (vii) Contribution to AI Literacy. The Embodied Understanding of AI occurs because this method uses physical role-playing and card-based games to simulate core AI mechanisms—such as neural networks, decision theory, machine learning, and pattern recognition—making abstract concepts accessible through movement, collaboration, and tangible materials. This method is designed for Non-Experts because the activities are designed to be implemented by elementary school teachers with no background in AI or programming, enabling AI literacy without requiring technical expertise. This opens AI education to a broader audience than ever before.The Cross-disciplinary Integration arises for the reason that these games link AI with mathematics, music, physical education, ethics, and language arts, allowing seamless integration into the regular curriculum. This interdisciplinary model is rarely seen in AI education at any level, and almost never at the elementary level. Why this method is Scientifically Grounded ? The answer is that each activity is built upon core scientific principles of AI and learning theory, and connects explicitly to foundational models such as feedforward neural networks, probabilistic decision-making, and classification algorithms. The educational design is backed by current literature in AI literacy, computational thinking, and embodied cognition. Here, we analyze the Original Game Designs: To the best of our knowledge, no previous work has proposed or implemented these specific physical game designs for teaching AI to children. These classroom games are novel contributions to both AI education and educational game design. This works talks about Open Educational Resource Potential due to the fact that this work lays the foundation for a future AI Education Game Kit—an open-access resource that can be distributed globally, especially in under-resourced schools where technology access is limited but physical teaching tools can be easily deployed.The contribution to AI can be justified because this paper supports recent frameworks of AI literacy for children, but moves beyond theory by providing tested, concrete, hands-on implementations that are aligned with modern pedagogy, making it a strong candidate for publication in AI in Education journals or presentations at AI conferences.

\section{Method}

The proposed pedagogical framework consists of four distinct physical games that target core AI concepts, specifically designed for primary school classrooms. Each game engages participants in embodied roles that simulate fundamental AI components or algorithms, fostering experiential learning. The suite is complemented by an integrative, cross-disciplinary project named 'Classroom Spotify', which synthesizes these concepts in a collaborative environment. The specifications of each game consist of (i) the objective of each game, (ii) the materials that are used, (iii) the guidelines to play the game, and (iv) the tips for teachers. The games mentioned are indicative, but represent a first approach to the method. The interaction between teachers and students is the key to the success of this new method. 
\subsection{The Classroom Neural Network (CNN) }
\subsubsection{Objective}
This game is designed to concretize functioning artificial neurons by demonstrating weighted input summation and threshold activation. The main aim of this game is to simulate the way in which a neural neuron receives the signals, analyses them, and finally makes a decision or prediction. The students will be able to realize the way that a machine learns from its mistakes.
\subsubsection{Materials}
The materials for this game are : 
\begin{itemize}
    \item[(i)] coloured ropes with different weights .
    \item[(ii)] Different t-shirts with the same  number on the back and the front.
    \item[(iii)] A colored T-shirt.
    \item[(iv)]  Cards of signals. 
\end{itemize}
\subsubsection{Instructions}

Each of the students who wants to participate in this game is a neuron. The participants are connected with different colored ropes with different weights. The students realize that these ropes are different either by looking at the color or checking the different weight. Each neuron receives the signals from its neighbors, and after multiplying it with the weight of connections, the neuron sums the results. The activation value for each neuron is arbitrarily defined, and it is what we call threshold. The threshold is notated by the number on the t-shirt. If the sum is larger than the threshold, then the neuron is activated and the student-neuron raises the hand. This movement of raising the hand is signal 1, and if a student does not raise the hand, this means that the signal is 0. At the end, one or more neurons are the final neurons that will give the prediction. If this prediction is incorrect, then the model needs corrections by changing the weights. Assume that this little classroom neural network wants to decide if the input signal of a red color is correct. The teacher selects 5 students as neurons of this little classroom neural network and 1 student as the user who will ask the system if the color is red. One of the students wears a red t-shirt. This student is the input neuron that receives the signal. The other 4 students (neurons) wear different t-shirts with numbers on the back and on the front. As we defined before, each of these numbers corresponds to the neuron's threshold activation level. Activation of a neuron is simulated by raising the hand. More precisely, the student who wears the red t-shirt raises a hand when the teacher selects a card with a red color. The other students who wear the t-shirts with the numbers raise their hands when they are active neurons. Let $r_{1},r_{2},r_{3},...r_{n}$ be the colored ropes with weights $1,2,3,...,n$ respectively. The 5 neurons are the input neuron $R_{i}$, the three neurons $B_{2},C_{2},D_{2}$ have threshold 2, and neuron $E_{3}$ has threshold 3. Assume that the classroom neural network is $CNN=\{R_{i}r_{1}B_{2},R_{i}r_{2}C_{2},B_{2}r_{1}D_{2},C_{2}r_{1}D_{2},D_{2}r_{3}E_{3}\}$. The input neuron $R_{i}$ is connected with its neighbors $B_{2}$ and $C_{2}$ with ropes of weights 1 and 2, respectively. The neurons $B_{2}$ and $C_{2}$ are connected with the neighbor $D_{2}$ using a rope of weight 1, but they are not connected to each other. Finally, the neuron $D_{2}$ has a connection of weight 3 with the neuron $E_{3}$. The student-user raises up the card with the red color, which corresponds to the initial signal for the CNN. Then, the student, who is the neuron $R_{i}$, raises his hand (1). The neuron $B_{2}$ receives a signal 1x1=1. This value is less than the defined threshold of 2, so the student $B_{2}$ does not raise up his hand (0). The neuron $C_{2}$ receives a signal of 1x2=2, so the threshold bound is satisfied, and the student $C_{2}$ raises his hand (1). To continue with, the total received signal for the neuron $D_{2}$ is 0x1+1x1=1, so the student $D_{2}$ does not raise his hand (0). The final decision of CNN is strongly connected with the signal that arises to the neuron $E_{3}$. This signal is equal to 0, so the student $E_{3}$ will not raise his hand (0). This means that the network's decision is negative about the red color of the card. It is evident that the decision was wrong, and it is crucial for the stability of the network to detect and fix the problems. For example, if we change the ropes, the total energy to the final neurons might be increased. 
\subsubsection{Reflexion}
“The Classroom Neural Network (CNN)” game offers a vivid and physical representation of core concepts underlying artificial neural networks, including weighted inputs, threshold-based activation, signal propagation, and error correction. Through embodiment and role-playing, students experience firsthand how interconnected nodes (neurons) process information based on input stimuli and internal thresholds. This kinesthetic approach demystifies the “black box” nature of machine learning and fosters intuitive understanding of abstract computational mechanisms. In addition, the game highlights the importance of internal parameters, such as weights and thresholds, in determining the behavior and performance of the network. Students can observe how even small changes to the weights (represented by colored ropes) significantly alter the final output, mirroring the process of training in artificial intelligence. The iterative refinement of weights following incorrect predictions introduces the foundational concept of learning from error—an essential aspect of supervised learning algorithms. This playful yet rigorous simulation aligns closely with educational neuroscience and constructivist pedagogies, where learning is rooted in active participation and sensory engagement. In addition, the collaborative nature of the activity fosters teamwork, critical thinking, and reflection on cause-and-effect relationships within AI systems. As a result, 'The Classroom Neural Network' serves not only as a didactic tool to understand technical principles but also as an accessible entry point into the broader discourse of machine learning and computational reasoning.

\subsection{The Surprise Box }
\subsubsection{Objective}
The Surprise box game is an interactive application of decision-making theory. The aim of this game is to simulate a model of possible choices, uncertainty, analysis of the relation between cost and benefit, and finally, the value of information. The students observe in practice how an information affects the final preference on a choice problem.
\subsubsection{Materials}
The materials for this game are :
\begin{itemize}
    \item[(i)] 2 boxes or more
    \item[(ii)] Award cards or surprise gifts
    \item[(iii)] A piece of paper sheet
    \item[(iv)] Information cards with costs.
\end{itemize}
\subsubsection{Instructions}
The students decide to open either box A or box B. There are 2 awards with respect to the amount of points. The first award corresponds to 100 points, and the second award corresponds to 30 points. The award cards are placed inside the boxes, and it is evident that neither the students nor the teachers know the content of each box. There are different cards with information about the content of the boxes. Moreover, these cards have a cost of reducing the total points of each choice. The participants decide if they want to purchase a random information card or not. Let $A_{c}^{i}$ be the card A with cost c and i\% the probability of finding the award of 100 points. Assume that $P_{A}(S)=\{ A_{20}^{10},B_{30}^{20},C_{5}^{5},D_{85}^{40}\}$ is the set of information cards for box A and $P_{B}(S)=\{ E_{10}^{50},F_{10}^{10},G_{20}^{30},H_{5}^{5}\}$ the set of information cards for box B. The teacher creates two groups of 4 students, one from each box. The teacher is the participant who does not make any choice. One student records the final results of points for each student. In small elementary school classes, the probability can be interpreted as the difficulty of finding the award of highest points.
\subsubsection{Reflexion} 
“The Surprise Box” game introduces students to fundamental principles of decision theory, including uncertainty, expected value, cost-benefit analysis, and the informational value of probabilistic data. By framing these abstract concepts within a tangible and playful context, the activity enables young learners to engage deeply with the logic behind decision-making processes. The use of probabilistic information cards with associated costs encourages students to weigh the trade-off between acquiring information and maximizing rewards, thus simulating real-world scenarios where decisions must be made under incomplete knowledge. Importantly, the game fosters an intuitive understanding of how information, even when uncertain, can alter rational preferences and influence outcomes. It also subtly introduces the idea that not all information is equally valuable, a notion central to both economic reasoning and artificial intelligence. The collaborative and competitive nature of the activity promotes active engagement, while the randomness inherent in the setup maintains excitement and motivation. In general, 'The Surprise Box' offers a rich educational experience, effectively translating complex theoretical ideas into experiential learning and laying the foundation for a more advanced exploration of decision-making systems and AI ethics.
\subsection{The Little trainers }
\subsubsection{Objective}
This game is designed to simulate the procedure of training a system via natural game. 
\subsubsection{Materials}
The materials for this game are :
\begin{itemize}
    \item[(i)] Data cards like animals, objects or simple shapes.
    \item[(ii)] Label cards like "dog" , "apple", etc.
    \item[(iii)] Question cards
    \item[(iv)] Feedback cards "YES" and "NOT"
    \item[(v)] A cardboard or corkboard with push pins
\end{itemize}
\subsubsection{Instructions}
The teacher assigns the roles to the students. The roles are organized into four basic groups of students, simulating training procedures (Group A), classification (Group B), testing (Group C)and feedback (Group D). Some students play the role of the trainer of this system. This group A presents the data of the system by showing the label for each class. For example, assume that we have different species of dogs and cats. The students present the cards with the animals organized in classes by showing the label for each class, hence using the card "DOG" and the card "CAT." Another team of students simulates the model, which classifies the data on the basis of the labels given. The third team is the tester team, which adds new data and asks the model to predict the class of the new item. The answer of the model will be evaluated by the referee team, which is responsible for crucial feedback. More specifically, assume that Group A has made a collection of 3 cards with dogs and 3 cards with cats from different species. This is the initial system that the students want to increase its stability to correct predictions and answers. The group learns to classify the objects, taking into account the features of the animals such as shapes, sounds, images, etc. Group C brings a new card with the image of a wolf, the features of which are similar to the features of a dog, asking the model to make a prediction about the label of this new object. The system may recognize this object as "DOG," maybe not. In case the system replies "DOG," Group D gives the crucial feedback that the answer was wrong. To fix this problem, the little trainers will assign the label of "WOLF" to this new object and repeat the same procedure again and again. (algorithm).
\subsubsection{Reflexion}
This activity provides a compelling and age-appropriate simulation of the supervised learning process in artificial intelligence. By assigning distinct roles to student groups—trainers, classifiers, testers, and feedback providers—it transforms abstract computational concepts into concrete social interactions. The division of labor reflects the real-world pipeline of machine learning, where data labeling, model training, testing, and error correction are interdependent and iterative. Through this structured role-play, students begin to grasp key AI concepts such as classification accuracy, generalization, overfitting, and the significance of feedback in refining models. The scenario involving the classification of a wolf as a dog illustrates the inherent ambiguity and limitations of pattern recognition systems when exposed to unfamiliar data. This moment of 'misclassification' offers a powerful learning opportunity, reinforcing the importance of diverse training data and continuous system refinement. Moreover, the collaborative nature of the game fosters communication, critical thinking, and teamwork—skills that are as vital in AI development as they are in educational settings. Overall, "The Little Trainers" effectively bridges cognitive and computational thinking, laying a strong pedagogical foundation for understanding intelligent systems.
\subsection{The predictors}

\subsubsection{Objective}
The students learn the fundamental concept of the pattern, the pattern of the data, the recognition of the pattern, and finally the prediction based on a pattern. The concept of training also exists in this game. The students observe that there is no specific amount of data that is enough for a good prediction.
\subsubsection{Materials}
The materials for this game are :
\begin{itemize}
    \item[(i)] Data cards with trigonometric shapes, colors, animals, etc.
    \item[(ii)] Data cards with numbers
    \item[(iii)] Data cards with images 
    \item[(iv)] A cardboard or corkboard with push pins
\end{itemize}
\subsubsection{Instructions}

The teacher presents a sequence of cards pressed in a cardboard or cork board. This sequence of cards maybe describes a pattern that the students should recognize in order to continue by predicting the next objects in the sequence. The challenge here is that despite the fact that this sequence might be evident, the teacher can change the pattern to underline the importance of the size of data. For example, assume that the teacher has pinned the cards with the symbols \MVAt,\Smiley,\textdollar in the pattern P=\{\MVAt,\Smiley,\textdollar,\MVAt,\Smiley,\textdollar,.....\}. If he asks the students about the next object, the answer is \MVAt. This is the moment where the challenge starts. The first scenario is that this answer is correct, so the student will become happy by making their first prediction. On the contrary, the second scenario is that this answer is false by giving the correct database  P=\{\MVAt,\Smiley,\textdollar,\MVAt,\Smiley,\textdollar,1,2,3,\MVAt,\Smiley,\textdollar,\MVAt,\Smiley,\textdollar,1,2,3,.....\}. This will be a great surprise for the students, and it will spark their interest in this game. The imagination and inspiration of the teacher in creating new sequences is the key to success for this activity.
\subsubsection{Reflexion}
This activity elegantly demonstrates the challenges and surprises inherent in data interpretation and pattern recognition—core elements of artificial intelligence and machine learning. By introducing seemingly simple sequences that later reveal deeper, more complex structures, students experience firsthand the limitations of small data samples and the dangers of premature generalization. The shift from an apparent pattern to a more nuanced one encourages learners to adopt a more critical and inquisitive stance, promoting habits of mind essential for data-driven reasoning. Moreover, the use of tangible materials and visual stimuli enhances engagement and supports diverse learning styles. Importantly, the teacher’s role as a designer of evolving sequences plays a pivotal part in maintaining curiosity and sustaining cognitive engagement. This hands-on experience fosters a foundational understanding of the importance of dataset size, hidden variables, and the iterative nature of model refinement—principles that mirror real-world challenges in AI development.

\subsection{The Classroom Spotify Project}

\subsubsection{Objective}
This project combines all the knowledge gained from the 4 basic games that we discussed before. This educational activity is designed to offer the experience of constructing an artificial intelligence machine, simulating a modern music platform like Spotify. The students will have the opportunity to construct their own smart music machine, which listens to the samples of the songs, recognizes the features like rhythm, lyrics, and style, and suggests a song to the user according to their preferences.  
\subsubsection{Materials}
The materials for this game are :
\begin{itemize}
    \item[(i)] Data cards with songs with description (title,icon,rhythm,emotion)
    \item[(ii)] Data cards with characteristics (relaxing,intense,party)
    \item[(iii)] board for scores
    \item[(iv)] board for feedback
    \item[(v)] board for RLID system of classification
\end{itemize}
\subsubsection{Instructions}

One of the students plays the role of the user who expresses his mood, for example, "I am feeling sleepy," "I am excited," or "I feel lonely." A group of 2 or more students selects a song, and another group of 2 or more plays the role of sensors. The sensors analyze the song as far as the rhythm (R), the subject of lyrics (L), the instruments (I), and the danceability (D) are concerned. The classification of the song with respect to these 4 classes is based on a rating scale from 1 to 3. This is the RLID system of rating for the sensors. A higher score is interpreted as a more positive sign for each characteristic of the RLID system. For example, if the sensors classify a song as RLID = (3, 3, 3, 3), this means that the song has a very fast tempo, very cheerful lyrics, very loud instruments, and it is very danceable. On the contrary, if the sensors classify a song as RLID = (1, 1, 1, 1), this means that the song has a very slow tempo, very sad lyrics, and very soft instruments, and it is not danceable at all. The sensors pin the classification information to a board. A group of students plays the role of the neurons. The neurons combine all the information from the sensors and make the scores for each song. For example, a song with classification (1,1,1,1) is labeled as song 4 from the neurons. The board for scores now contains the coded information for the "decider" group. The decider group observes the scores on the board and suggests a song to the user according to this preference. The feedback student asks the user if this song satisfies his mood and records the response. If his answer is "YES," then this is one of the appropriate songs for his mood; hence, the feedback student pins the card of this song in the feedback board. It's crucial for the feedback board to be organized according to the different moods of the users. If the answer is "NO," then the student who is responsible for the feedback keeps notes about the reason for this negative response. The Classroom Spotify Machine will be updated every time we repeat the same procedure for all the students and for all their different moods.   
\subsubsection{Reflexion}
The Classroom Spotify Project serves as a capstone educational experience, integrating key artificial intelligence (AI) concepts introduced in previous games—signal processing, feature recognition, decision theory, neural activation, and feedback loops—into a holistic simulation of an intelligent music recommendation system. Inspired by real-world platforms like Spotify, this project invites students to embody the roles of sensors, classifiers, and decision-making agents, thereby experiencing firsthand how AI models operate in real-life applications. The RLID classification system (Rhythm, Lyrics, Instruments, Danceability) provides a concrete framework for the students to engage in feature extraction and multidimensional evaluation of data. This mirrors the preprocessing phase in machine learning pipelines where raw inputs are transformed into structured representations. The students acting as "neurons" and "deciders" replicate the internal logic of recommender systems, making decisions based on aggregated data while accounting for subjective user preferences. A particularly powerful pedagogical component is the incorporation of feedback. The feedback mechanism mimics the reinforcement learning paradigm, where system performance is iteratively refined through user responses. In this context, the evolving “feedback board” becomes an analogue to the dynamic memory of AI models, reinforcing associations between input features and user satisfaction. By simulating the entire lifecycle of a recommendation engine—from data collection and classification to decision-making and adaptation—this project fosters systems thinking, interdisciplinary learning, and ethical reflection on algorithmic influence. Furthermore, it demonstrates how AI technologies are embedded in everyday experiences, such as music consumption, making abstract computational processes accessible and meaningful to young learners. This comprehensive and experiential approach cultivates AI literacy in an age-appropriate and deeply engaging manner.

\section{Discussion}
As Generation Alpha children are growing up immersed in AI-powered environments—from voice assistants and recommendation algorithms to educational apps, the need for foundational AI literacy in primary education becomes increasingly urgent. Studies have highlighted the importance of early exposure to AI concepts to foster critical thinking, ethical awareness, and informed citizenship in the digital age \cite{tour}. Without structured guidance, children risk becoming passive users of intelligent systems, unable to question or understand the mechanisms that shape their digital experiences. As argued by Williams et al. (2021) \cite{will}, AI education should begin in primary school through age-appropriate, hands-on activities that demystify machine learning, data bias, and algorithmic decision-making. Integrating AI fundamentals into early curricula empowers students to engage actively and responsibly with the technology that increasingly influences their lives. The physical and tangible approach to AI education presented here addresses critical gaps in early AI literacy by providing accessible, engaging, and conceptually rich learning experiences. By concretizing abstract AI processes through role-play and manipulatives, these games lower cognitive barriers and foster intuitive understanding. The Classroom Spotify project exemplifies how individual concepts can be integrated into complex systems, promoting systems thinking and collaborative skills. Future work should empirically evaluate learning outcomes across diverse populations and explore digital augmentations to enhance scalability. This approach may also be adapted for cross-disciplinary STEM integration, expanding its educational impact. This innovative framework is a new tool for the elementary teachers, to plan some courses on the basis of teaching fundamental concepts of artificial intelligence machines. This work provides the motivation for conducting empirical and practical research on the results of these applications in children's knowledge of computer science. It is evident, that there is a great interaction between students and teachers who don't have any knowledge of AI before. The AI machines provide detailed instructions to use them as a teaching assistant tool. On the other hand, there are not so many applications to the core of learning AI. The query that arises now is, has the time come to revise the curricula for elementary school students?The answer will be provided through research that must continue not only at a theoretical level with the creation of interactive applications like those proposed, but also at a practical level with the assessment of learning through these games.

\par

\par\bigskip\smallskip\par\noindent

\par\noindent
{\it Address}: {\tt {Nikolaos Sampanis} \\ {Department of Computer Engineering \& Informatics\\ University of Patras\\ Greece}}
\par\noindent
{\it E-mail address}:{\tt nsampanis@upatras.gr}

\end{document}